\documentclass{aa}
\usepackage{graphicx}
\usepackage{natbib}
\usepackage{amssymb}
\bibpunct{(}{)}{;}{a}{}{,}
\begin{document}
   \title{Statistical properties of exoplanets}

   \subtitle{III. Planet properties and stellar multiplicity}
   
   \author{A. Eggenberger \and S. Udry \and M. Mayor}

   \offprints{Anne Eggenberger, \email{Anne.Eggenberger@obs.unige.ch}}

   \institute{Observatoire de Gen\`eve, 51 ch. des Maillettes, 
             CH-1290 Sauverny, Switzerland
             }

   \date{Received / Accepted  }

   \abstract{ Among the hundred or so extrasolar planets discovered to date, 
   19 are orbiting a component of a double or multiple star system. 
   In this paper, we discuss
   the properties of these planets and compare them to the characteristics 
   of planets orbiting isolated stars. Although the 
   sample of planets found in multiple star systems is not large, some
   differences between the orbital parameters and the masses of 
   these planets and the ones of planets orbiting single stars are emerging 
   in the mass--period and in the eccentricity--period 
   diagrams. As pointed out by \cite{Zucker02b}, the most massive
   short-period planets are all found in multiple star systems. We show here
   that the planets orbiting in multiple star systems also tend to have
   a very low eccentricity when their period is shorter than about 40 days. 
   These observations seem to indicate that some kind of migration has been at
   work in the history of these systems. The properties of the five 
   short-period planets orbiting in multiple star systems seem, however, 
   difficult to explain with the current models of planet formation and 
   evolution, at least if we want to invoke a single mechanism to account for 
   all the characteristics of these planets.
   
   }

   \titlerunning{Statistical properties of exoplanets III}

   \maketitle

\defcitealias{Udry03}{Paper~I}


\section{Introduction}

Studies of stellar multiplicity among solar-type stars of the solar
neighbourhood have shown that about 40\% of the G and K dwarfs can be
considered to be real single stars \citep{Duquennoy91,Eggenberger03c}. 
As the majority of solar-type stars belong to double or multiple star 
systems\footnote{In this paper, double and multiple star systems 
will be called multiple star systems.}, it is
of interest to consider the existence of planets in such an environment. 
Searches for extrasolar planets using the radial velocity technique have shown
that giant planets exist in certain types of multiple star systems (see Table 1 for further details). The number of
such planets is, however, still low, in part because close binaries are difficult 
targets for radial velocity surveys and were consequently often rejected from
the samples. Due to the limitations of the available observational techniques, 
most detected objects are giant (Jupiter-like) planets; the existence of 
smaller mass planets in multiple star systems is thus still an open question.

\begin{table*}

\caption{Planets orbiting a component of a multiple star system with confirmed
orbital or common proper motion (CPM stands for common proper motion 
and SB for spectroscopic binary).}
\begin{center}
\begin{tabular}{lcccccc}
\hline
Star & a$_{{\rm b}}$ & a$_{{\rm p}}$ & M$_{{\rm p}}\sin{i}$ & e$_{{\rm p}}$ & Notes & References\\
     &  (AU)           &           (AU) &  (M$_{\rm J}$)       &                &       &           \\
\hline
\object{HD\,40979}              & $\sim$6400        & 0.811    & 3.32   & 0.23  & CPM$^{a}$& 12,11  \\
\object{Gl\,777\,A}             & $\sim$3000        & 4.8     & 1.33   & 0.48   & CPM           & 1,23 \\
\object{HD\,80606}              & $\sim$1200        & 0.469   & 3.90   & 0.927 & CPM            & 22  \\
\object{55\,Cnc}                & $\sim$1065        & 0.115   & 0.84   & 0.02  & CPM            & 8,21,19,2  \\
                                &                   & 0.24   & 0.21   & 0.34  &                &          \\
                                &                   & 5.9     & 4.05   & 0.16  &                &          \\
\object{16\,Cyg\,B}             & $\sim$850         & 1.6    & 1.5   & 0.634  & CPM            & 24,5,15  \\
\object{Ups\,And}               & $\sim$750         & 0.059   & 0.71   & 0.034 & CPM            & 17,24,2,3\\
                                &                   & 0.83   & 2.11   & 0.18  &                &  \\
                                &                   & 2.50    & 4.61  & 0.44  &                &  \\
\object{HD\,178911\,B}          & $\sim$640         & 0.32    & 6.292  & 0.1243& CPM            & 28,30  \\
\object{HD\,219542\,B}          & $\sim$288         & 0.46    & 0.30   & 0.32  & CPM            & 7 \\
\object{Tau\,Boo}               & $\sim$240         & 0.05    & 4.08   & 0.018  & orbit          & 13,24,2\\
\object{HD\,195019}             & $\sim$150         & 0.14    & 3.51   & 0.03   & CPM            & 24,1,10 \\
\object{HD\,114762}             & $\sim$130         & 0.35    & 11.03  & 0.34  & CPM            & 24,16,18  \\
\object{HD\,19994}              & $\sim$100         & 1.54     & 1.78   & 0.33   & orbit          & 13,25,20 \\
\object{HD\,41004\,A}           & $\sim$23          & 1.33    & 2.5    & 0.39  & SB             & 29,31,27 \\
\object{$\gamma$\,Cep}          & $\sim$22          & 2.03    & 1.59   & 0.2   & SB             & 4,6,14 \\
\object{Gl\,86}                 & $\sim$20          & 0.11    & 4.0    & 0.046 & CPM, SB$^{b}$ & 9,26 \\
\hline
\end{tabular}
\end{center}
\label{tab1}
{\small { Notes: (a) According to \citet{Halbwachs86}, this pair has only a
probability of 60\% to be physical. The physical nature of this binary has
however been confirmed later on the basis of 
CORAVEL radial velocity measurements (Halbwachs, private communication); }
(b) The multiplicity status of this system has still to be clarified.

\vspace{1.5mm}

 References: (1) \citealt{Allen00}; (2) \citealt{Butler97}; (3) \citealt{Butler99};
(4) \citealt{Campbell88}; 
(5) \citealt{Cochran97}; (6) \citealt{Cochran02}; (7) \citealt{Desidera03};
(8) \citealt{Duquennoy91}; (9) \citealt{Els01}; 
(10) \citealt{Fischer99}; 
(11) \citealt{Fischer03}; 
(12) \citealt{Halbwachs86}; (13) \citealt{Hale94}; 
(14) \citealt{Hatzes03};
(15) \citealt{Hauser99}; 
(16) \citealt{Latham89}; (17) \citealt{Lowrance02}; 
(18) \citealt{Marcy99}; (19) \citealt{Marcy02}; (20) \citealt{Mayor03}; 
(21) \citealt{McGrath02}; 
(22) \citealt{Naef01}; (23) \citealt{Naef03}; (24) \citealt{Patience02};  
(25) \citealt{Queloz00b}; (26) \citealt{Queloz00}; (27) \citealt{Santos02}; 
(28) \citealt{Tokovinin00}; (29) \citealt{Udry03spec}; 
(30) \citealt{Zucker02}; (31) \citealt{Zucker03b}.}
\end{table*}

The orbital characteristics and the mass distribution of extrasolar planets 
can give us an insight into their formation mechanisms and their subsequent
evolution. In the first paper of this series, 
\citeauthor{Udry03} (\citeyear{Udry03}; \citetalias{Udry03}) 
discussed the period distribution and 
the mass--period diagram for extrasolar planets orbiting single stars. 
As pointed out by \citet{Zucker02b}, planets orbiting a component of a 
multiple star system seem to have different characteristics than planets 
orbiting single stars. \citet{Zucker02b} showed that there is a significant
correlation in the mass--period diagram for planets orbiting single stars, 
while there may be an anticorrelation in this same diagram for planets found in 
multiple star systems. The difference is mainly due to a paucity of 
massive planets with short periods, and to the fact that the most massive 
short-period planets are all found in binaries.

The characteristics of extrasolar giant planets have forced considerable
modifications of the standard model of planet formation. It is now usually
believed that planets form within a protoplanetary disc of gas and dust orbiting
a central star, but the precise modes by which this formation takes place are
still debated, especially for giant planets 
\citep[e.g.][]{Pollack96,Boss97,Bodenheimer00,Boss00b,Wuchterl00,Boss03}. 
Two major models have been proposed to explain giant planet formation 
(see Sect.~\ref{giantformation}), each with its 
advantages and limitations, but there is currently no consistent model that 
accounts for all the observed characteristics of extrasolar planets. Observational
constraints are thus needed, not only to specify our understanding of
planet formation and evolution, but also to possibly discriminate between the
proposed models. In this context, the detection and the characterization
of planets orbiting in multiple star systems, even if more difficult to carry out 
than the study of planets orbiting isolated stars, may bring 
new constraints and additional information.

This paper is organized as follows. 
The sample of planets found in multiple star systems is presented in 
Sect.~\ref{obs}.  
Some trends seen in the statistics are then emphasized in
Sect.~\ref{stat}. 
Models of formation and evolution of giant planets in binaries are briefly
reviewed in Sect.~\ref{models} and their predictions are 
compared to the observations in Sect.~\ref{discussion}. Our conclusions 
are drawn in Sect.~\ref{conclusion}.

\section{Known planets in multiple star systems}
\label{obs}

Among the extrasolar planets discovered to date, some of them are orbiting a
component of a multiple star system. Planets have been found around stars known to
be part of a wide common proper motion pair, known to be in a visual binary or
in a spectroscopic binary. 
Alternatively, searches for faint companions to stars hosting planets 
have revealed a few new systems. These observations, summarized 
in Table 1, show that giant planets can form and survive in certain types of 
multiple star systems.

\section{Statistics of planets in multiple star systems}
\label{stat}

Although the sample of planets found in multiple star systems is not large,
a preliminary comparison between the characteristics of these planets and 
the ones of planets orbiting isolated stars can be made. 
Here, we will discuss the mass--period and the eccentricity--period diagrams for
extrasolar planets, focusing on possible differences between the two
populations. Our sample of planetary candidates orbiting a component  
of a multiple star 
system consists of all the systems listed in Table~\ref{tab1}. The total sample of
extrasolar planetary candidates is made of 115 objects\footnote{see e.g. 
http://obswww.unige.ch/Exoplanets/} with a minimum mass 
$M_2\sin{i}\leq 18$\,M$_{\mathrm J}$. The orbital elements used for the 
analysis are
the ones deduced from our radial velocity data or the most recent version of 
the ones given in the literature.

\subsection{The mass--period diagram}
\label{mp_diag}

\begin{figure}
\resizebox{\hsize}{!}{\includegraphics{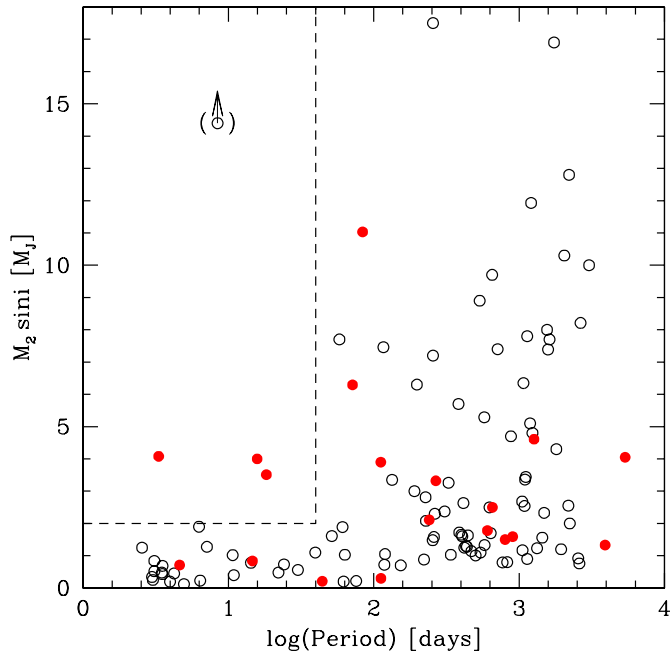}}
\caption{Planetary minimum mass versus orbital period for the known extrasolar 
planetary candidates. Planets orbiting a single star are represented by open
circles, 
while planets orbiting a component of a multiple star system are represented by
filled circles. The dashed line approximately delimits the zone where 
only extrasolar planets belonging to multiple star systems are found.}
\label{diag_m_logp}
\end{figure}

Figure~\ref{diag_m_logp} shows the distribution of all the extrasolar 
planetary candidates
in the $M_2\sin{i}$--$\log{P}$ plane. Two interesting features emerge 
from this plot: there are no short-period extrasolar 
planets with a mass $M_2\sin{i} \gtrsim 5$\,M$_{\mathrm J}$, and 
the most massive short-period planets are almost 
all found in multiple star systems (\citealp{Udry02a,Zucker02b};
\citetalias{Udry03}). Indeed, planetary candidates with a mass 
$M_2\sin{i} \gtrsim 2$\,M$_{\mathrm J}$ and a period $P \lesssim 40$ days are 
all orbiting a component of a multiple star system, the only exception 
being \object{HD\,162020\,b}. As explained in \citet{Udry02a}, HD\,162020\,b is 
probably a brown dwarf with a true mass much larger than its minimum mass, 
and should therefore be removed from our diagram. If it is true that no planet 
with a mass $M_2\sin{i} \gtrsim 2$\,M$_{\mathrm J}$ orbiting a single star has 
been found with a period $P \lesssim 40$ days, the two populations of
planets are somewhat mixed together for periods between 40 and 150 days. 
The orbital period below which there is no more massive planet orbiting a 
single star is thus not well defined, and such a unique 
and well defined limit may, in fact, not exist.

\begin{figure}
\resizebox{\hsize}{!}{\includegraphics{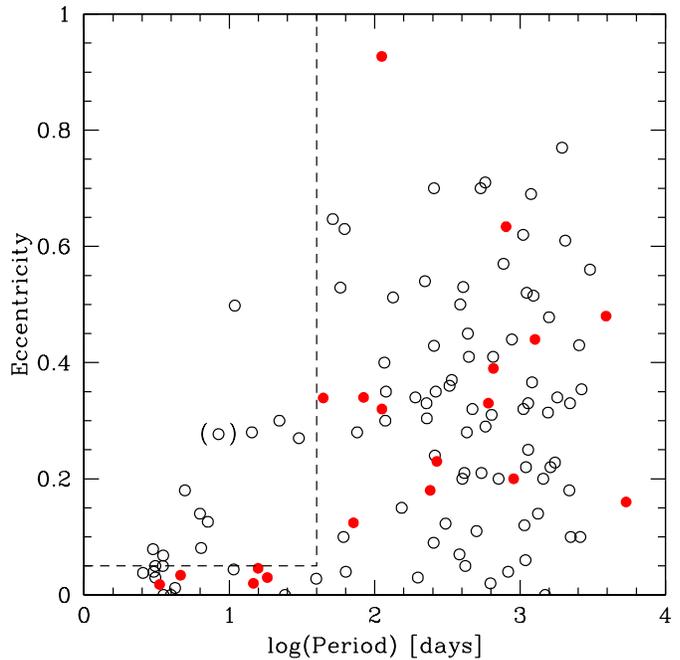}}
\caption{Eccentricity versus orbital period for all the extrasolar planetary
candidates. Same symbol coding as in Fig.~\ref{diag_m_logp}. The dashed line 
approximately delimits the zone where there are no planets belonging to 
multiple star systems.}
\label{diag_e_logp}
\end{figure}

The paucity of massive short-period planets cannot be attributed to 
observational selection effects since these planets are the
easiest to detect. Moreover, even if the sample of planets orbiting a component
of a multiple star system is small and incomplete, the presence of a
few candidates in a zone of the diagram where there are no other planets is
significant. We will come back to these differences in Sect.~\ref{discussion}.

For periods longer than $\sim$100 days, the distribution of the planetary
companions in the $M_2\sin{i}$--$\log{P}$ plane is not very different for the
two samples. In this period range, the mean mass of planets found 
in multiple star systems is, nevertheless, smaller than the mean 
mass of planets
orbiting single stars. This difference comes from the fact that no
very massive planet has been found on a long-period orbit around a component 
of a multiple star system. Again, this cannot be attributed to 
observational selection effects, because several planets with a smaller mass 
and on long-period orbits have been found in multiple
star systems. 

The lack of planets with $M_2\sin{i} \geq 5$\,M$_{\rm J}$ and 
$P \geq 100$ days in multiple star systems can, however, be due to the 
small number of planets found in multiple star systems. 
To check this point, we computed the probability
that, drawing a random subsample of 10 planets (i.e. the number of long-period
planets found in multiple star systems when selection effects are taken into
account) with $P \geq 100$ days out of the total 
population of planets with $P \geq 100$ days, we would 
have no planet with $M_2\sin{i} \geq 5$\,M$_{\rm J}$ (see Appendix \ref{a_mlogp}
for further details). Using the hypergeometric
distribution (Appendix \ref{hypergeo}), this probability is $3.25$\,\% 
(selection effects have been taken into account by discarding from the counts 
the planets with a radial velocity semiamplitude $K < 15$\,ms$^{-1}$). 
Thus, if the planets with $P \geq 100$ days orbiting in multiple star 
systems have the same properties as the planets with $P \geq 100$ days 
orbiting single stars, the probability not to have a single planet with 
$P \geq 100$ days and $M_2\sin{i} \geq 5$\,M$_{\rm J}$ among our sample of
planets found in multiple star systems is $3.25$\,\%. This result 
shows that this trend could be real. It is, however, not possible to 
exclude with a high confidence level that the lack of 
planets with $M_2\sin{i} \geq 5$\,M$_{\rm J}$ and $P \geq 100$ days may solely 
be due to small-number statistics and that planets with $P \geq 100$ days
orbiting in multiple star systems and around isolated stars may, in fact, belong
to the same population. A larger sample of long-period planets orbiting 
in multiple star systems will be required to settle this question.

\subsection{The eccentricity--period diagram}
\label{ep_diag}

The distribution of the extrasolar planetary candidates in the e--$\log{P}$
plane is illustrated in Fig.~\ref{diag_e_logp}. In this diagram, we note 
that all the planets with a period $P \lesssim 40$ days orbiting in 
multiple star systems have an eccentricity smaller than 0.05, whereas  
longer period planets found in multiple star systems can 
have larger eccentricities. 
Some of the very short-period planets are so close to their parent star that 
tidal dissipation in the planet could have circularized their orbit, 
even if they were originally eccentric \citep{Rasio96}. 
For longer periods, the orbits are not necessarily circularized anymore and
any eccentricity is possible. For the three planets with a period between 10 and
40 days orbiting in multiple star systems, the circularization time (due to
tidal dissipation in the planet) is $\tau_{c}\gtrsim 10^{12}$ years. 
This is clearly too long to explain the low eccentricities of these planets by
invoking tidal dissipation alone.

The hypergeometric distribution (Appendix \ref{hypergeo}) 
was again used to test if the difference
observed in this diagram may solely be due to the small size of the sample of
short-period planets found in multiple star systems. In this case, we computed 
the probability to have 5 planets with $\log{P} \leq 1.6$ and $e<0.05$ in a
subsample of 5 planets with $\log{P} \leq 1.6$ drawn out of the total 
population of planets with $\log{P} \leq 1.6$ (see Appendix \ref{a_elogp}
for more details). When selection effects are taken into account 
(by discarding from the counts the planets with a radial velocity semiamplitude 
$K < 15$\,ms$^{-1}$), this probability is $3.77$\,\%. 
This means that the probability to have 5 planets 
with $\log{P} \leq 1.6$ and $e<0.05$ in the sample of short-period planets
orbiting in multiple star systems is $3.77$\,\% if these planets have the same
properties as short-period planets orbiting isolated stars. 
On the basis of the current samples, there is thus a trend towards a 
difference in the properties of short-period planets orbiting in multiple star 
systems and around isolated stars. The alternative statement, namely that 
short-period 
planets orbiting in multiple star systems and around isolated stars belong to 
the same population can, nevertheless, not be excluded. 
The observation that the three extrasolar planets
with periods between 10 and 40 days orbiting a component of a multiple star 
system have very low eccentricities is interesting and could be a clue to their 
formation and/or subsequent evolution history. We will come back to this point 
in Sect.~\ref{discussion}.

For periods longer than $\sim$40 days, there is no significant 
difference between 
planets found in multiple star systems and around isolated stars. The
mean eccentricity is similar for the two samples.

\subsection{Remarks}
\label{rem}

The limit at $\sim$40 days is valid for both the 
eccentricity--period and the mass--period diagrams, but in the latter the 
limiting period is less well defined. If we plot the
evolution of the mean mass or the highest mass (averaged on the three highest
values) of planets orbiting single stars as a function of the period 
(see Fig.~6 of \citetalias{Udry03}), there is a jump
at a period of $\sim$40 days, which reflects the distribution of these planets
in the mass--period diagram. Thus, about the
same limiting period is obtained by considering the distribution of planets
orbiting single stars in the mass--period diagram on the one hand, and by
considering the distribution of planets orbiting in multiple star systems in
the eccentricity--period diagram on the other hand. This is intriguing.

Even if the orbital parameters of the binaries hosting planets are not exactly
known, the projected separations of these systems (see Table 1) 
indicate that the five planets with a period shorter that 40 days reside in
very different types of systems. There is thus no obvious correlation between
the properties of these planets and the known orbital characteristics 
of the binaries or the star masses.

The history of planets found in multiple 
planet systems is probably different from the one of "single" planets. 
For the analysis presented here, all planets have been considered, 
but we have checked that our
conclusions remain unchanged if the planets belonging to multiple 
planet systems are removed from the samples. Among the five short-period 
planets found in multiple star systems, two also belong to multiple planet
systems. This is something that we must keep in mind when discussing the
properties of these planets.

\section{Models of formation and evolution of giant planets in binaries}
\label{models}

Let us now turn to planet formation models and
consider their predictions regarding the existence and the survival of giant
planets in binary star systems. We will then be in position to
compare and confront our observations with their results.

There are different points to take into account when considering the
formation of giant planets in binaries. Indeed, the stellar companion
affects all the stages of planet formation as well as the subsequent
evolution of the planet once it has formed. The major points that have
been studied in the literature are briefly described in this section.

\subsection{Binary-disc interactions}
\label{bin_disc}

Binary stars can in principle interact with three types of discs: two
circumstellar and one circumbinary. Transfer of angular momentum between the
binary and the disc leads to a truncation of the inner/outer edge for a 
circumbinary/circumstellar disc, respectively.  

\citet{Artymowicz94} investigated the approximate sizes of discs as a 
function of binary mass ratio and eccentricity for systems with 
circumstellar and circumbinary gaseous discs. 
Their results show that for a
binary with a mass parameter $\mu=M_2/(M_1+M_2)=0.3$, the inner
edge of a circumbinary disc is typically located at  
$r_{\mathrm t}=2.0\,a_{\rm b}$ (where $a_{\rm b}$ is the binary semimajor axis) 
for nearly circular binaries and at 
$r_{\mathrm t}=3.0\,a_{\rm b}$ for $e_{\rm b}=0.5$. 
The outer edge of a circumprimary disc lies at 
$r_{\mathrm t}=0.4\,a_{\rm b}$ for nearly circular binaries and 
at $r_{\mathrm t}=0.18\,a_{\rm b}$ for $e_{\rm b}=0.5$.
For a circumsecondary disc, the outer truncation radius is located near  
$r_{\mathrm t}=0.27\,a_{\rm b}$ for nearly circular binaries and 
near $r_{\mathrm t}=0.15\,a_{\rm b}$ for $e_{\rm b}=0.4$.

\subsection{Long-term stability of orbits}
\label{stability}

Assuming that planets can form in binary stars, do long-term stability
regions exist for planetary orbits in these systems? 
\citet{Holman99} studied the long-time survival of planets in different regions
of phase space near a binary star system. Circumprimary
as well as circumbinary planets were studied for different values of the binary
eccentricity and mass ratio. For a binary with a mass parameter $\mu=0.3$,
the largest stable orbit around the primary star has a critical semimajor axis 
$r_{\mathrm c}=0.37\,a_{\rm b}$ for $e_{\rm b}=0.0$ or 
$r_{\mathrm c}=0.14\,a_{\rm b}$ for $e_{\rm b}=0.5$. For the same binary, the
smallest circumbinary orbit has a critical semimajor axis  
$r_{\mathrm c}=2.3\,a_{\rm b}$ for $e_{\rm b}=0.0$ or 
$r_{\mathrm c}=3.9\,a_{\rm b}$ for $e_{\rm b}=0.5$.

The study of \citet{Holman99} was restricted to planets in initially circular 
motion. 
\citet{pilat01} extended this type of analysis and determined the 
variation of the stable zone due to an increase of the initial planet
eccentricity. An increase of the
planet eccentricity reduces the stable zone, this reduction being of course less
pronounced that the one due to the same increase of the binary
eccentricity.

As shown by \citet{Holman99} a companion star orbiting beyond 
about 5 times the
planetary distance is not a serious threat to the long-term stability of 
planetary orbits. Nevertheless, this result only applies to orbits with a 
low mutual inclination. \citet{Innanen97} investigated the stability of 
planetary orbits in binary systems
with emphasis on the inclination of the orbital planes. As an example, they
studied the stability of the solar system under the presence of an hypothetical
distant companion placed at 400\,AU with different inclinations and masses. 
Due to the Kozai
mechanism \citep{Kozai62}, the system is unstable at high inclination when the companion 
mass is larger than 0.05\,M$_{\sun}$. A low-mass companion does, however, not 
destabilize the system, even when the inclination is high.

\subsection{Giant planet formation in binaries}
\label{giantformation}

Two mechanisms have been proposed to explain giant planet formation: core 
accretion and
disc instability \citep[e.g.][]{Pollack96,Boss97,Mayer02}. 
The core accretion mechanism begins with the collisional
accumulation of planetesimals and planetary embryos in a protoplanetary disc, 
as for terrestrial planet formation. In the outer parts of the disc, where the 
amount of solid material is increased by the presence of ices, embryos may 
reach about 10\,M$_{\oplus}$ in $\sim$$10^6$ years and 
begin to grow an atmosphere of disc gas to form giant planets like Jupiter and
Saturn in $\sim$$10^7$\,years.  
In the disc instability model, a gravitationally unstable
disc fragments directly into self-gravitating clumps of gas and dust that can
contract and become giant gaseous protoplanets. Coagulation and sedimentation 
of dust grains to the protoplanet center could form a solid core. 
This process occurs over a dynamical time scale: clump formation and dust grain 
sedimentation proceed nearly simultaneously in $\sim$$10^3$\,years. 
Though they are different, these two mechanisms share a common 
characteristic: giant planets should only form in the relatively cool outer 
regions of protoplanetary discs.

\citet{Boss98} considered the influence of a binary companion on giant planet 
formation via disc instability. 3D hydrodynamical models of discs with 
0.04\,M$_{\sun}$ were evolved
in time, subject to the gravity of a binary star companion placed on a circular orbit
at 40\,AU. In the absence of the binary companion, the disc is stable, but in
the presence of the binary companion the disc forms a multi-Jupiter-mass
protoplanet in 0.002\,Myr.
 
The evolution of two stars, each orbited by a circumstellar disc, 
was simulated by \citet{Nelson00} using a two
dimensional smoothed particle hydrodynamic code. Each component of the binary
had a mass of 0.5\,M$_{\sun}$ and a binary eccentricity of 0.3 was considered.
The system was evolved over 2700\,yr (8 binary orbits). During and 
after periastron each disc developed strong two-armed spiral structures
which 
decayed to a smooth condition over the next half binary period; 
this cycle repeating with little variations. The spiral structures decay was due
to internal heating in the discs, which increased their stability against
spiral arm growth. Giant planet formation via gravitational 
collapse is therefore unlikely in this system. In fact, the temperatures 
in the discs are so high, that some grain species, including water ices, 
are vaporized everywhere. Giant planet formation by 
the core accretion mechanism is thus unlikely as well in this system.

\subsection{Evolution of an embedded planet in a binary}
\label{embedded}

A different approach was considered by \citet{Kley00} who studied the evolution of
a giant planet still embedded in a protoplanetary disc around the primary
component of a binary system. A 1\,M$_{\rm{J}}$ planet was placed on a circular
orbit at 5.2\,AU from a 1\,M$_{\sun}$ star. The secondary star had a mass of
0.5\,M$_{\sun}$ and an eccentricity of 0.5. The binary semimajor axis was varied
from 50 to 100 AU. The simulations show that the companion alters the 
evolutionary properties of the planet: the mass accretion rate
is increased and the inward migration time is reduced.

\subsection{Summary}

The main effect a companion has on a protoplanetary disc is a 
truncation of 
its radius and an induction of waves which, upon dissipation, 
transfer angular momentum between the binary and the disc. Comparing
the results presented in Sect.~\ref{bin_disc} and in Sect.~\ref{stability}, 
we see that a
planet can almost always persists for a long time, wherever it forms in 
a truncated protoplanetary disc. 
The effect a secondary star has on the efficiency of planet formation is, 
however, less clear. 
According to \citet{Nelson00}, the companion has a negative influence, 
slowing or inhibiting altogether giant planet formation. \citet{Boss98} 
claims the opposite, namely that giant 
planet formation via gravitational collapse is favoured in binaries. 
More comprehensive studies will be needed, not only to clear up the case of
binaries with a separation of 40 or 50\,AU, but also 
to explore what happens for binaries with
different projected separations, mass ratios and eccentricities. 
Anyway, we expect the secondary star
to have an influence on planet formation, at least for close binaries, and
\citet{Kley00} has shown that this remains true for  
the subsequent evolution of giant planets in binaries. Planets found in 
multiple stars systems may thus
have different characteristics than planets orbiting isolated stars.

\section{Discussion}
\label{discussion}

As mentioned in Sect.~\ref{giantformation}, in situ formation is very 
unlikely for short-period Jupiter-mass planets. Formation at larger distances 
followed by inward migration seems to be a better explanation to the 
existence of these planets \citep[e.g.][]{Lin96,Bodenheimer00}. The high masses
and low eccentricities of short-period planets orbiting in multiple star 
systems (Sect.~\ref{obs} and Figs.~\ref{diag_m_logp} and \ref{diag_e_logp}) 
seem also to indicate that some kind of migration has been at work in the
history of these systems. A few different migration mechanisms have been proposed
to explain the existence and the characteristics of short-period giant planets.
We will now briefly discuss two of them and see if they might explain some
of the features emphasized in Sect.~\ref{obs}.

\subsection{Planet--viscous disc interaction}

One proposed migration mechanism involves the gravitational interaction 
of a protoplanet with the gaseous disc out of which it formed 
\citep{Goldreich79,Goldreich80,Ward86,Ward97}. 
Subject to such an interaction, high mass planets will migrate 
more slowly than low mass planets in a 
given disc because they create larger gaps. 
Moreover, some of these planets will experience mass
loss as they come close to the central star. In overall, we thus expect 
to find more massive planets at intermediate and large semimajor axes, 
the population of close-in objects being dominated by 
smaller mass planets \citep{Trilling02}. This is indeed observed
\citepalias{Udry03} and \citet{Zucker02b} have
shown that this effect is statistically significant.

Now, if we consider the evolution of a protoplanet orbiting the primary star of
a binary system, we have seen that the presence of the companion alters the
evolutionary properties of the planet, in particular the migration and mass
growth rates are enhanced \citep{Kley00}. These differences may explain
why the most massive short-period planets are found in multiple star systems:
either they are more massive because of the higher mass accretion rate, or they
are massive planets like the ones found with periods longer than $\sim$100 
days around single stars, but orbiting closer-in    
in multiple star systems because of the higher migration rate. Both of these
effects are in fact probably mixed together and present at the same time.

Still regarding migration via the gravitational interaction of a Jupiter-mass
protoplanet with a gaseous disc, models indicate that a realistic upper
limit for the masses of closely orbiting giant planets is $\sim$5\,M$_{\rm J}$, 
if they originate in protoplanetary discs similar to the minimum-mass
solar nebula \citep{NelsonR00}. Examples of large 
($> 5$\,M$_{\rm J}$) planets at small orbital distances can, however, 
be obtained due to migration in discs with different masses or viscosities 
\citep{Trilling98}. 
If such an upper mass limit exists for short-period
planets, and if the scenario proposed by \citet{Kley00} is correct, we would
expect to find an upper mass limit for short-period planets orbiting in 
multiple star systems that is larger than the one valid for planets orbiting 
single stars.
Therefore, there should exist a zone in the mass--period diagram where only
planets orbiting in multiple star systems are found, the presence of a stellar
companion being the reason that enables a planet to reach a small separation 
with a mass larger than the limit corresponding to planets orbiting 
isolated stars.

In the simulations by \citet{Kley00}, the planet eccentricity is also
modified: it first grows due to the perturbations induced by the secondary star,
but then declines because of the damping action of the disc. The final result is
a rapid decay of the planet semimajor axis and a damping of the initial
eccentricity. 

Taken at face value, these arguments may provide an explanation for the 
observation that the most massive short-period planets are all found in 
multiple 
star systems and have very small eccentricities. It should, however, be noticed that several of 
the multiple star systems known to host planets are probably very different 
from the ones studied by \citet{Kley00}. The five planets with a period 
shorter than 40 days orbit in binaries with very different separations 
(from $\sim$20 to $\sim$1000\,AU) and it seems not likely that the perturbations
produced by a wide companion would influence the evolution of a protoplanet
orbiting at or below a few AU. This, however, deserves further study.

\subsection{Kozai migration}

Another mechanism that may be at work in wide binaries is the so-called
Kozai migration \citep{Wu03b,Wu03}. In such a case, the Kozai mechanism 
(\citealp{Kozai62}; see also \citealp{Holman97,Innanen97,Mazeh97}) 
produces large cyclic oscillations of the planet eccentricity. 
During the periods of high eccentricity, the periastron is
small and, consequently, tidal dissipation becomes important and 
gradually removes energy from the planetary orbit, eventually leading to 
circularization. The Kozai mechanism coupled with tidal dissipation is thus 
a viable method by which a planet can migrate towards the central star.

Tidal dissipation depends sensitively on the nearest approach distance and is 
important only if the planet can reach a high eccentricity. As the eccentricity
oscillations only depend on the inclination of the planet orbital plane 
relative to the binary orbital plane, the initial inclination between 
these two planes is a key parameter. Moreover, for the Kozai mechanism to be 
effective, the companion must provide the 
dominant contribution to the apsidal precession of the planet (see
\citealp{Holman97} and \citealp{Wu03} for more details). 
The efficiency of the Kozai migration is, however, fairly independent of the 
planet mass, and this mechanism will work for planets of relatively large 
masses \citep{Wu03b}.

Although the Kozai mechanism may be efficient in binaries with large
semimajor axes, several requirements must be simultaneously satisfied 
for it to operate, and such a mechanism
will not apply to a large fraction of planetary systems. 
Furthermore, even if 
the Kozai migration has been efficient during a period in the evolution of a
planet, it does not imply that its orbit is now circular. 
Up to now, the Kozai mechanism and the Kozai migration have been 
considered to 
explain the high eccentricity of given planetary candidates such as 
\object{16\,Cyg\,B\,b} and \object{HD\,80606\,b} 
\citep{Holman97,Mazeh97,Wu03}. 
It has never been demonstrated that the combination 
of the Kozai mechanism with tidal dissipation may account for the existence of
close-in planets with very low eccentricities. 
On the other hand, Kozai oscillations are not likely to be currently at work 
for the planets with short semimajor axes, in particular 
because the Kozai mechanism is suppressed by general 
relativistic effects.
It is thus  very unlikely that the low eccentricity of these planets may be due
to the fact that they are currently seen in their low-eccentricity phase. 
It seems therefore difficult to explain the characteristics of all 
the short-period planets orbiting in multiple stars systems by invoking 
the Kozai migration alone.

\section{Conclusion}
\label{conclusion}

The characteristics of giant planets 
found in multiple star systems seem to be different from the ones of 
planets 
orbiting single stars, at least for the short-period planets. The
major differences are:
\begin{itemize}
\item the most massive ($M_2\sin{i}\gtrsim 2$\,M$_{\mathrm J}$) 
short-period planets all orbit in multiple star systems;
\item the planets found in multiple star systems tend to have a very low 
eccentricity when their period is shorter than 40 days.
\end{itemize}
These observations seem to indicate that migration has played an important role
in the history of the short-period planets orbiting in multiple star systems and
that migration may be induced differently in binaries than around 
single stars. 

From the theoretical point of view, it has been shown \citep{Kley00} that 
the presence of a companion star affects the properties 
of a Jupiter-mass planet still embedded in a
disc around the primary component of a binary by increasing the migration
and mass accretion rates. Alternatively, the Kozai mechanism may
be (or have been) at work in binaries hosting planets and 
will modify some of the orbital 
parameters of the planet. This mechanism can be efficient in wide binaries and, 
coupled with tidal dissipation, it may 
also lead to inward migration. Even if these 
two mechanisms may be invoked to explain the characteristics 
of a few planets orbiting in binaries, none of them seem to be able to
account for all the properties of the five planets orbiting in multiple stars
systems with a period $P\lesssim 40$ days. Nonetheless, it is also possible 
that diverse mechanisms may have been at work in
these systems, but leading to a similar final state and similar planet
properties. New studies dedicated to this issue will be needed to settle 
this question and to find a satisfactory explanation to 
the existence and the characteristics of the short-period planets 
found in multiple star systems.

From the observational point of view, a larger sample of planets orbiting in
multiple star systems will be required to confirm or refute the preliminary 
trends emphasized in this paper. In this context, the search for planets in
multiple star systems, even if more difficult to carry out than the search for
planets around single stars is of importance. On the other hand, 
the characterization 
of the star systems susceptible of hosting planets is underway and could
bring interesting constraints for the models, thus helping our understanding 
of giant planet formation.

\appendix
\section{The hypergeometric distribution}
\label{hypergeo}

The hypergeometric distribution models the total number of successes in a 
fixed size sample drawn without replacement from a finite population of $N$ items
of which $G$ are labelled {\sl success} and $(N-G)$ are labelled {\sl failure}. The
hypergeometric distribution is described by three parameters: 
$N$, the size of the population; $G$, the total number of items with the 
desired characteristics in the
population; and $n$, the size of the random sample drawn from the population. 

The probability distribution of the hypergeometric random variable $X$, 
the number of successes in a random sample of size $n$ selected from the total 
population is:

$$P(X=x) = \frac{C_x^G C_{n-x}^{N-G}}{C_n^N}$$

where $x = \mathrm{max}(0,n-(N-G)),\ldots,\mathrm{min}(n,G)$. 
This probability formula represents the ratio of the number of samples 
containing $x$ successes and $(n-x)$ failures to the total number of possible 
samples of size $n$.

To test the statistical significance of the possible difference observed in 
the mass--period or in the eccentricity--period diagram, we considered a 
subsample of planets drawn from the total population of extrasolar planets 
(i.e. planets found in multiple star systems and around isolated stars). 
A planet was labelled {\sl success} if it was located within the test zone of 
the diagram considered and {\sl failure} otherwise. We then computed the 
probability that such a random subsample would give rise to a similar 
configuration as the one actually observed for planets found in multiple 
star systems, namely a configuration with the same number of planets within 
the test zone.

\subsection{The mass--period diagram}
\label{a_mlogp}

We give here more details regarding the 
statistical significance of the lack of planets
with $M_2\sin{i}\geq 5$\,M$_{\rm J}$ and $P \geq 100$ days 
orbiting in multiple star systems 
(Fig.~\ref{diag_m_logp} and Sect.~\ref{mp_diag}). As explained 
in Sect.~\ref{mp_diag}, the hypergeometric distribution was used to compute 
the statistical significance of the difference observed. The
parameters used were: $N=77$, the number of planets with $P \geq 100$ days and
$K\ge15.0$\,ms$^{-1}$; 
$G=21$, the number of planets with $P \geq 100$ days, 
$M_2\sin{i} \geq 5$\,M$_{\rm J}$ and
$K\ge15.0$\,ms$^{-1}$; $n=10$, 
the number of planets with $P \geq 100$ days and
$K\ge15.0$\,ms$^{-1}$ found in multiple star systems; and $x=0$, 
the number of planets with $P \geq 100$ days, $K\ge15.0$\,ms$^{-1}$ and 
$M_2\sin{i} \geq 5$\,M$_{\rm J}$ 
found in multiple star systems. Given these parameters,
we obtain a probability $P(X=0)= 3.25$\,\% to have no planet with 
$P \geq 100$ days, $K\ge15.0$\,ms$^{-1}$ and $M_2\sin{i} \geq 5$\,M$_{\rm J}$ 
among a subsample of 10 planets drawn from the total population of planets.

\subsection{The eccentricity--period diagram}
\label{a_elogp}

More details concerning the statistical significance of the possible difference
observed for short-period planets in the eccentricity--period diagram 
(Fig. \ref{diag_e_logp} and Sect.~\ref{ep_diag}) are given here. 
The parameters used to compute the
hypergeometric probability were: $N=25$, the number of planets with 
$\log{P} \leq 1.6$ and $K\ge15.0$\,ms$^{-1}$; 
$G=14$, the number of planets with 
$\log{P} \leq 1.6$, $K\ge15.0$\,ms$^{-1}$ and $e<0.05$; 
$n=5$, the number of planets with 
$\log{P} \leq 1.6$ and $K\ge15.0$\,ms$^{-1}$ orbiting in multiple star systems; 
and $x=5$, the number of planets with 
$\log{P} \leq 1.6$, $K\ge15.0$\,ms$^{-1}$  and $e<0.05$ found in
multiple star systems. The probability to have 5 planets with 
$\log{P} \leq 1.6$, $K\ge15.0$\,ms$^{-1}$ and $e<0.05$ in a subsample of 5
planets drawn from the total population of planets is then 
$P(X=5)= 3.77$\,\%.

\begin{acknowledgements}
We thank the Swiss National Research Foundation (FNRS) and the Geneva 
University for their continuous support to our planet search programmes. 
We thank the anonymous referee for valuable suggestions regarding the
statistics. This
research has made use of the SIMBAD database and the VizieR catalogue access
tool operated at CDS, France. 
\end{acknowledgements}


\bibliographystyle{aa} 
\bibliography{0164bib}

\end{document}